\documentclass[reqno]{amsart}
\usepackage{times,mathrsfs,amsmath,amssymb,amsfonts}
\usepackage{latexsym,textcomp,verbatim}
\usepackage[hypertex,linkcolor=red]{hyperref}
\newtheorem{definition}{Definition}
\newtheorem{lemma}{Lemma}

\newtheorem{example}{Example}
\def\eg{e. g. }

\def\set#1{{\sf #1}}
\def\sH{\set{H}}\def\sK{\set{K}}

\def\<{\langle}\def\>{\rangle}
\def\Bnd#1{\set{B(#1)}}
\def\Tr{\operatorname{Tr}}\def\d{\operatorname{d}}

\def\map#1{{\mathscr{#1}}}
\def\Proof{\medskip\par\noindent{\bf Proof. }} 
\def\dual#1{{#1}^\tau}

\def\Reals{\mathbb R}
\def\Index#1{{\em #1}}

\def\:{{\mathbf :}}
\def\Klm{{\mathfrak X}}
\def\ltwo{{\mathbf l}_2}\def\Ltwo{{\mathbf L}_2}
\def\Reals{\mathbb R}\def\Cmplx{\mathbb C}
\def\kk{\rangle\!\rangle}\def\bb{\<\!\<}
\newcommand{\dk}[1]{| \, #1 \rangle\!\rangle}
\newcommand{\db}[1]{\<\!\< #1 \, |}  
\begin{document}
\title{Renormalized quantum tomography} \author[Giacomo Mauro D'Ariano]{Giacomo Mauro D'Ariano}
\email{dariano@unipv.it} \address{{\em QUIT} Group, Dipartimento di Fisica ``A. Volta'' and INFN
  Sezione di Pavia, via A. Bassi 6, 27100 Pavia, Italy\\ and\\
  Center for Photonic Communication and Computing, Department of Electrical and Computer
  Engineering, Northwestern University, Evanston, IL 60208-3118, USA} \author[Massimiliano Federico
Sacchi]{Massimiliano Federico Sacchi}\email{msacchi@unipv.it} \address{{\em QUIT} Group,
CNR-INFM and CNISM Unit\`a di Pavia, via A. Bassi 6, I-27100 Pavia, Italy}
\begin{abstract}
  The core of quantum tomography is the possibility of writing a generally unbounded complex
  operator in form of an expansion over operators that are generally nonlinear functions of a
  generally continuous set of spectral densities----the so-called {\em quorum} of observables. The
  expansion is generally non unique, the non unicity allowing further optimization for given
  criteria.  The mathematical problem of tomography is thus the classification of all such operator
  expansions for given (suitably closed) linear spaces of unbounded operators---e.g. Banach spaces
  of operators with an appropriate norm.  Such problem is a difficult one, and remains still open,
  involving the theory of general basis in Banach spaces, a still unfinished chapter of analysis. In
  this paper we present new nontrivial operator expansions for the quorum of quadratures of the
  harmonic oscillator, and introduce a first very preliminary general framework to generate new
  expansions based on the Kolmogorov construction. The material presented in this paper is intended
  to be helpful for the solution of the general problem of quantum tomography in infinite
  dimensions, which corresponds to provide a coherent mathematical framework for operator expansions
  over functions of a continuous set of spectral densities.
\end{abstract}
\maketitle
\tableofcontents
\thispagestyle{empty}
\section{Introduction}
\label{sec:intro}

The state of a physical system is the mathematical description that provides a complete information
on the system. Its knowledge is equivalent to know the result of any possible measurement on the
system. In classical mechanics it is always possible, at least in principle, to devise a procedure
made of multiple measurements which fully recovers the state of a single system. In quantum
mechanics, on the contrary, there is no way, not even in principle, to infer the quantum state of a
single system without having some prior knowledge on it \cite{imposs}. It is however possible to
estimate the quantum state of a system when many identical copies are available prepared in the same
state, so that a different measurement can be performed on each copy.  Such a procedure is called
{\em quantum tomography}.  

\par The problem of finding a strategy for determining the state of a system from multiple copies
dates back to 1957 by Fano \cite{fano}, who called {\em quorum} a set of observables sufficient for
a complete determination of the density matrix.  However, quantum tomography entered the realm of
experiments more recently, with the pioneering experiments by Raymer's group \cite{raymer} in the
domain of quantum optics. In quantum optics, in fact, using a balanced homodyne detector one has the
unique opportunity of measuring all possible linear combinations of position and momentum---the
so-called {\em quadratures}---of the harmonic oscillator representing a single mode of the radiation
field.

\par The first technique to reconstruct the density matrix from homodyne measurements --- so called
{\em homodyne tomography} --- originated from the observation by Vogel and Risken \cite{vogel} that
the collection of probability distributions achieved by homodyne detection is just the Radon
transform of the Wigner function $W$. Therefore, similarly to classical imaging, one can obtain $W$
by inverting the Radon transform, and then from $W$ one can recover the matrix elements of the
density operator.  This original method, however, works well only in a semi-classical regime, whereas
generally for small photon numbers it is affected by an unknown bias caused by the smoothing
procedure needed for the analytical inversion of the Radon transform. The solution to the problem is
to bypass the evaluation of the Wigner function, and to evaluate the matrix elements of the density
operator by simply averaging suitable functions of homodyne data: this is the basis of the first
unbiased topographic technique presented in Ref. \cite{dmp}. Clearly the state is perfectly
recovered in principle only in the limit of infinitely many measurements: however, for finitely many
measurements one can estimate the statistical error affecting each matrix element.  For infinite
dimensions there is the further problem that the propagation of statistical errors of the density
matrix elements make them useless for estimating the ensemble average of some operators (\eg
unbounded), and a method for estimating the ensemble average is needed, which bypasses the
evaluation of the density matrix itself, as was first suggested in Ref.  \cite{tokio}. For a brief historical
excursus on quantum tomography, along with a review on the generalization to any number of radiation
modes, arbitrary quantum systems, noise deconvolution, adaptive methods, and maximum-likelihood
strategies the reader is addressed to Ref. \cite{revt}.

\par The most comprehensive theoretical approach to quantum tomography uses the concept of {\em
  frame of observables}, i.~e. a set of observables spanning the linear space of operators, from
which one derives contextually the {\em quorum} of observables and the estimation rule. The ensemble
average $\< H\>$ of any arbitrary operator $H$ on a Hilbert space $\sH$ is estimated using
measurement outcomes of the quorum $\{ O_l\}$ upon expanding $H$ over a set of functions $f_n(O_l)$
of the observables $\{ O_l\}$. What makes the general theory nontrivial in infinite dimensions is
the crucial role of the non linear functions $f_n(O_l)$ in making the infinite expansion convergent.
Let's denote by $P_j:=f_n(O_l)$, $j=(n,l)$, such complete set of operators.  Once you have the
$P_j$, then the problem is reduced to the {\em linear} problem of expanding an operator as $H=\sum_j
\< Q_j^\dag, H\> P_j$, for a suitable ``dual'' set of operators $\{Q_j\}$. (Notice that generally
the index $l$ is continuous, whence also the operator expansion.) The scalar product in the
expansion is generally not simply the Frobenius's when we need to expand operators that are not
Hilbert-Schmidt.  The mathematical {\em theory of frames} \cite{Christiansen, Casazza,Han,Li} is the
perfect tool for establishing completeness of $\{P_j\}$ and for finding dual sets $\{Q_j\}$. In most
practical situations the set $\{P_j\}$ is over-complete, and there are many alternate dual sets
$\{Q_j\}$, the non unicity providing room for optimization.  A general theory should also classify
the operators $H$ having bounded scalar product with $\{Q_l\}$ and expansion $H=\sum_j \< Q_j^\dag,
H\> P_j$ converging in average for a given class of quantum states.  In infinite dimensions---e.~g.
for homodyne tomography---the easy known approach works only for Hilbert-Schmidt operators, whence
including trace-class ones---e.~g. for estimating the matrix elements of the density operator over an
orthonormal basis. For unbounded operators, however, such operator expansion becomes an infinite sum
of unbounded terms.  On the other hand, converging (and even finite) alternate expansions are known
to exist for various unbounded operators \cite{Rich}.  As we will see in this paper, the mechanism
allowing "renormalization" of the expansion relies on the existence of {\em
  null-estimators}---namely operator-valued functions that have zero mean over the quorum---their
existence being related to a group of symmetries of the quorum. The notion of null-estimator was
first introduced in Ref.  \cite{null}, in the context of of homodyne tomography, where the
quorum is made of the quadratures of the field mode---the quantum analogue of the Radon transform.
Here the symmetry group of the quorum is the group $U(1)$ of rotations of the quadrature phase. The
existence of null-estimators leads to infinitely many alternate expansions of the same operator over
the quorum, allowing cancellations of the infinities in the expansion---a kind of "renormalization"
procedure.

The problem of classifying all operator expansions for a given quorum in infinite dimensions for
given spaces of unbounded operators is a difficult one, and remains still open. It involves the
theory of {\em frames} or even more general notions of basis in Banach spaces \cite{Christiansen,
  Casazza,Han,Li}, a still unfinished chapter of analysis. In this paper we present new nontrivial
operator expansions for the quorum of quadratures of the harmonic oscillator, and introduce a first
preliminary general framework to generate and classify new expansions, based on the Kolmogorov
construction. We hope that the material presented here will open the way to the solution of the
problem of quantum tomography in infinite dimensions, leading to a general mathematical framework
for operator expansions over functions of a set of spectral densities.

\section{Quantum tomography and quorum of observables}
The general idea of quantum tomography is that there is a set of observables
$\{X_\xi\}_{\xi\in\Klm}$ 
on the Hilbert space of the system $\sH $---called "quorum"---by which one
can estimate any desired ensemble average by measuring the observables of the quorum, each at the
time, in a scheme of a repeated measurements. The observables of the quorum must be independent each
other, namely they are not commuting:
\begin{equation}
[X_{\xi'},X_{\xi''}]=0\;\Longleftrightarrow\; \xi'=\xi''.
\end{equation}
Generally the set $\Klm$ parameterizing the quorum is infinite, and most commonly, is a continuum. In
these cases, since clearly one can measure only a finite number of observables, these are randomly
picked out according to a given probability measure on $\Klm$, which, therefore, must be a
probability space. In the following, for simplicity, we will also assume a probability density over
$\Klm$ and denote it with the symbol $\d \mu(\xi)$. It follows that the ensemble average of a
(generally complex) operator is written in the form of double expectation
\begin{equation}
\< X\>=\int_\Klm\d\mu(\xi)\<f_\xi(X_\xi|X)\>,\label{doubleav}
\end{equation}
where the generally nonlinear function $f_\xi(x|X)$ of the variable $x$ has an analytic form which
depends on the particular operator $X$. We will call the function $f_\xi(x|X)$ the {\em tomographic
  estimator} for $X$ with quorum $\{X_\xi\}_{\xi\in\Klm}$. If we want to achieve the estimation of
$\< X\>=\Tr[\rho X]$---the expectation being supposedly bounded on the state $\rho$---by averaging
the estimator $f_\xi(x|X)$ over both quorum and measurement outcomes with a bounded variance, we
need to have the function $f_\xi(x|X)$ square-summable over $x$ and $\xi$, more precisely
\begin{equation}
\int_\Klm\d\mu(\xi)\int_{\mathcal{X}_\xi}\<\d E_\xi(x)\>|f_\xi(x|X)|^2 <\infty,\
\end{equation}
where $\mathcal{X}_\xi$ denotes the spectrum of $X_\xi$, and $\d E_\xi(x)$ its spectral measure. In
the following, for simplicity, we will consider the spectrum $\mathcal{X}_\xi\equiv\mathcal{X}$
independent on $\xi$. Clearly, the above square-summability will depend again on the state $\rho$
and on the operator $X$. 
\par We first want to notice two main features of estimators:
\begin{enumerate}
\item The estimator $f_\xi(x|X)$ is generally not unique, namely there can be many different
estimators for the same operator $X$.  This is equivalent to the existence of {\em null estimators},
namely functions $n_\xi(x)$ such that  
\begin{equation}
\int_\Klm\d\mu(\xi)\int_{\mathcal{X}}\d E_\xi(x)n_\xi(x)=0.
\end{equation}
Accordingly, the estimators are grouped into equivalence classes, each class corresponding to an
operator $X$. For such equivalence we will use the notation $\simeq$, i.~e. we will write $f\simeq
g$ or $f-g\simeq 0$ to denote that the two estimators are equivalent, namely they differ by a null estimator.
\item For fixed $x$ and $\xi$ the estimator $f_\xi(x|X)$ must be a linear functional of $X$,  namely
\begin{equation}
f_\xi(x|aX+bY)=af_\xi(x|X)+bf_\xi(x|Y),\qquad f_\xi(x|X^\dag)=f_\xi(x|X)^*.
\end{equation}
\end{enumerate}
\begin{example}[Homodyne tomography] \cite{revt} The quorum is given by $\{X_\phi\}_{[0,\pi)}$ with
  $X_\phi$ denoting the quadrature $X_\phi\doteq\tfrac{1}{2}(a^\dag e^{i\phi}+ae^{-i\phi})$, $a
  ,a^\dag $ being the annihilation and creation operators of the harmonic oscillator with commutator
  $[a, a^\dag ]=1$.  Estimators for the dyads $|n\>\< m|$ made with the orthonormal basis
  $\{|n\>\}$, $n=0,\ldots,\infty$ are
\begin{eqnarray}
f_\phi (x||n\>\< m|) &=&
\int_{-\infty}^{+\infty}\frac
{dk\,|k|}{4}\, e^{\frac{1-\eta}{8\eta}k^2-ikx}\<  n+d| e^{ik
X_{\phi}}| n\>  \label{estimat} \nonumber \\&= & 
e^{id(\phi+\frac{\pi}{2})}\sqrt{\frac{n!}{(n+d)!}}\int_{-\infty}^{+\infty}
dk\,|k| e^{\frac{1-2\eta}{2\eta}k^2-i2kx} k^d L_n^d(k^2)\;,
\end{eqnarray}
where $L_n^d(x)$ denotes the generalized Laguerre polynomials. 
\noindent For the unbounded operators $a$ and $a^\dag a$ one can check that the
following are unbiased estimators
\begin{eqnarray}
f_\phi (x|a)&=&2e^{i\phi}x, \nonumber \\
f_\phi (x|a^\dag a)&=&2x^2-\tfrac{1}{2}.
\;
\end{eqnarray}
\end{example}
\par The problem of quantum tomography is to establish the general rule for estimation, namely
\begin{definition}[Estimation rule] Given the quorum $\{X_\xi\}_{\xi\in\Klm}$ find the
  correspondence:
\begin{equation}
f_\xi(x|X)\Longleftrightarrow X,\qquad\text{for every operator}\, X
\text{on}\,\sH, 
\end{equation}
\end{definition}
where we possibly mean to find the whole equivalence class of estimators $f_\xi (X_\xi |X )$. 
\par Before solving this task, first one needs to know that the set of observables
$\{X_\xi\}_{\xi\in\Klm}$ is actually a quorum. The easiest thing to do, however, is to derive both
the quorum and the estimation rule contextually, starting from a spanning-set of
observables---shortly {\em observable spanning-set}---namely a set of observables
$\{F_\omega\}_{\omega\in\mathfrak{O}}$ in terms of which we can linearly expand operators as follows
\begin{equation}
X=\int_\mathfrak{O}\d\omega\; c_\omega(X) F_\omega.\label{obsframe}
\end{equation}
Notice that the notion of operator spanning-set used here generalizes the notion of frames for
Banach spaces to unbounded operators (see also the following), and is generally not strictly a frame
according to the definition of Refs. \cite{Christiansen,Casazza}.  In the following throughout the
paper we will always assume probability distributions admitting densities.  Generally, the set
$\mathfrak{O}$ is unbounded, and the measure $\d\omega$ is not normalizable, whence, as such, the
expansion (\ref{obsframe}) cannot be used for quantum tomography. However, generally this feature is
related to the redundancy of the observable spanning-set, which includes many observables $F_\omega$ that
are just different functions of the same observable. Then, collecting the observables of the spanning-set
into functional equivalence classes $\mathfrak{K}_\xi$, each corresponding to an observable of the
quorum $\{X_\xi\}_{\xi\in\Klm}$, one can relabel the observable spanning-set as $F_{\kappa,\xi}\doteq
f_{\kappa}(X_\xi)$ with $\kappa\in\mathfrak{K}_\xi$, writing
\begin{equation}
X=\int_\Klm\d\mu(\xi)\int_{\mathfrak{K}_\xi}\d\nu(\kappa)\; c_{\kappa,\xi}(X) f_{\kappa}(X_\xi)=
\int_\Klm\d\mu(\xi)f_{\xi}(X_\xi|X),
\end{equation}
where the function $f_{\xi}(x|X)$ is the integral over the observables
equivalent to $X_\xi$, namely
\begin{equation}
f_{\xi}(X_\xi|X)\doteq\int_{\mathfrak{K}_\xi}\d\nu(\kappa)\; c_{\kappa,\xi}(X) f_{\kappa}(X_\xi).
\end{equation}
Notice that in terms of the spectral decomposition of $X_\xi$ we can write 
\begin{equation}
 X=\int_\Klm\d\mu(\xi)\int_{\mathcal{X}_\xi}\d E_\xi(x) f_{\xi}(x|X),\label{XEx}
\end{equation}
and since this expansion is linear in the spectral measure $\d E_\xi(x)$, the latter can be regarded
itself as an observable spanning-set. Indeed, upon introducing the spectral density $\d
E_\xi(x)\doteq Z_{\xi,x}\d x$, and the density $\d\mu(\xi)\doteq m(\xi)\d\xi$, and renaming
$\zeta=(\xi,x)$ and $\mathfrak{Z}\doteq\{(\xi,x),x\in\mathcal{X}_\xi,\xi\in\Klm\}$ Eq. (\ref{XEx}) can
be rewritten in the same form of Eq. (\ref{obsframe}), namely
\begin{equation}
 X=\int_\mathfrak{Z}\d\zeta c'_\zeta(X) Z_\zeta, 
\end{equation}
where the new expansion coefficients are now given by 
\begin{equation}
c'_{\xi,x}(X)=m(\xi)f_\xi(x|X)=m(\xi)\int_{\mathfrak{K}_\xi}\d\nu(\kappa)\; c_{\kappa,\xi}(X) f_{\kappa}(x).
\label{cc}
\end{equation}
For homodyne tomography the above quantities are explicitly given in Table \ref{t:tomo}.
\par For $F_\omega $ an operator frame, the coefficients of the expansion (\ref{obsframe}) can be
rewritten in form of a pairing $(\cdot|\cdot)$ with a dual frame $G_\omega$ as follows
\begin{equation}
c_\omega(X)=(G_\omega|X),
\end{equation}
in terms of which Eq. (\ref{cc}) rewrites in the pairing form
\begin{equation}
c'_{\xi,x}(X)=m(\xi)(W_{\xi,x}|X),
\end{equation}
with dual frame
\begin{equation}
W_{\xi,x}=\int_{\mathfrak{K}_\xi}\d\nu(\kappa)\, G_{\xi,x}f_{\kappa}^*(x). 
\end{equation}
From the last equation it follows that the estimator itself can be written using the pairing
\begin{equation}
f_\xi(x|X)=(W_{\xi,x}|X),\label{pairingW}
\end{equation}
or, in terms of the original observable frame
\begin{equation}
f_\xi(x|X)=\int_{\mathfrak{K}_\xi}\d\nu(\kappa)\, (G_{\xi,x}|X) f_{\kappa}(x). 
\end{equation}
\begin{table}[hb]
\begin{center}
\begin{tabular}{|c|c|}
\hline
\hline
general & homodyne \\ 
\hline
\hline
$X_\xi$ & $X_\phi$ \\ 
\hline
$\Klm$ & $[0,\pi)$ \\ 
\hline
$\xi$ & $\phi$ \\ 
\hline
$\d\mu(\xi)$ & $\tfrac{\d\phi}{\pi} $ \\ 
\hline
$\omega$ &$\alpha $\\
\hline
$F_\omega$ &$D(\alpha)$\\
\hline
\hline
\end{tabular}
\begin{tabular}{|c|c|}
\hline
\hline
general & homodyne \\ 
\hline
\hline
$c_\omega(X)$ &$\Tr[D^\dag(\alpha)X]$\\
\hline
$\omega$ &$\alpha $\\
\hline
$\mathfrak{K}_\xi$ & ${\mathbb R}$ \\ 
\hline
$\d\nu(k)$ & $\frac{1}{4}\d k |k|$ \\ 
\hline
$c_{\xi,\kappa}(X)$ & $\Tr[e^{-ikX_\phi} X]$ \\ 
\hline
$f_\kappa(X_\xi|X)$ & $e^{ikX_\phi}$ \\ 
\hline
\hline
\end{tabular}
\begin{tabular}{|c|c|}
\hline
\hline
general & homodyne \\ 
\hline
\hline
$\d\mu(\xi)$ & $\tfrac{\d\phi}{\pi} $ \\ 
\hline
$Z_{\xi,x}$ &$|x\>_\phi{}_\phi\<x|$\\
\hline
$\mathcal{X}_\xi$ &${\mathbb R}$\\
\hline
$\mathfrak{Z}$ &$[0,\pi)\times{\mathbb R}$\\
\hline
$m(\xi)$ &$\frac{1}{\pi}$\\
\hline
$W_{\xi,x}$ & $-\frac{1}{4\pi}\frac{\operatorname{P}}{(x-X_\phi)^2}$ \\ 
\hline
\hline
\end{tabular}\end{center}
\caption{Table of correspondence for homodyne tomography.}\label{t:tomo}
\end{table}
\subsection{Unbiasing noise}
It is possible to estimate the {\em ideal} ensemble average $\<
X\>$ by measuring the quorum in the presence of instrumental noise, when the noise map $\map{N}$  
is invertible, or,  more generally, if there exists the right inverse of ${\map N}$. In terms of observable
frames this just corresponds to using a different dual frame. More precisely, one has: 
\begin{equation}
\begin{split}
X=\map{N}\map{N}^{-1}(X)=&\int_\Klm\d\mu(\xi) \map{N}[f_\xi(X_\xi|\map{N}^{-1}(X))]\\=&
\int_\Klm\d\mu(\xi)\int_{\mathcal{X}_\xi}\map{N}(\d E_\xi(x))
f_{\xi}(x|\map{N}^{-1}(X)). 
\end{split}
\label{unbias}
\end{equation}
Notice that Eq. (\ref{unbias}) means that
\begin{equation}
\<X\>=\int_\Klm\d\mu(\xi)\int_{\mathcal{X}_\xi}\<\d E_\xi(x)\>_\map{N} f_{\xi}(x|\map{N}^{-1}(X)),
\end{equation}
where $\<\,\cdot\,\>\doteq\Tr[\rho\,\cdot\,]$ denotes the ideal ensemble average, whereas
$\<\,\cdot\,\>_{\map{N}} \doteq\Tr[\dual{\map{N}}(\rho)\,\cdot\,]$ denotes the experimental ensemble
average, ${\map {N}}_*$ being the predual map of $\map {N}$
(Schroedinger 
versus Heisenberg picture).
This also means that for left invertible map $\map{N}$ the noisy spectral measures $\map{N}(\d
E_\xi(x))$ are still a quorum. In terms of the pairing in Eq. (\ref{pairingW}) unbiasing the noise
is equivalent to use the new dual frame $\map{N}^{-1}{}^\dag(W_{\xi,x})$.  When $\map{N}$ is not
right-invertible one can still estimate the ensemble average of operators in the range of the map.
Moreover, in infinite dimension, when the noise CP map $\map{N}$ is compact its inverse map is
unbounded, and one generally cannot unbias the noise without restricting the space of reconstructed
operators. Otherwise, one has a Hadamard ill-posed problem, for which there are biased compromises,
such as putting a cutoff on the vanishing singular values of $\map{N}$.
\begin{example}[Pauli tomography in a Pauli channel] \cite{revt} The operators
$[\sigma_\alpha/\sqrt2]$ make an observable orthonormal basis for
$\mathbb{C}^{\otimes 2}$. We consider now the noise described by the
depolarizing Pauli channel
\begin{equation}
{\map N}=(1-p)\map{I}+\frac p2 \map{T},
\end{equation}
where $\map{I}$ denotes the identity map and $\map{T}$ the trace map $\map{T}(X)\doteq I\Tr(X)$.
This noise can be simply unbiased via noise-map inversion: 
\begin{equation}
\map{N}^{-1\dag}=\frac{1}{1-p}\map{I}-\frac{p}{2(1-p)}\map{T}.
\end{equation}
\end{example}
\begin{example}[Homodyne tomography with $\eta< 1$] \cite{revt} The set of displacements operators
  $D(\alpha):=e^{\alpha a^\dag -\alpha ^* a}\,,\quad\alpha\in\mathbb C$ make an observable
  Dirac-orthonormal frame for $\set{T}_{1/2}$, where
\begin{equation}
\set{T}_s=\{X\in\set{T},\;
X=\: f(a,a^\dag)\:,\;\mbox{s.t.} \;
\lim_{\alpha\to\infty}f(\alpha,\bar{\alpha})e^{s|\alpha|^2}=0\}
\end{equation}
where $\: \ \ \:$ denotes normal ordering. 
In the presence of noise from nonunit quantum efficiency $\eta$,
the unbiased reconstruction is possible for operators in $\set{T}_s$
if $\eta\ge (2s)^{-1}$. In fact, one uses the new dual: 
\begin{equation}
D(\alpha)\to\map{N}^{-1\dag}(D(\alpha))=\eta
D(\eta^{1/2}\alpha)e^{\frac{1-\eta}{2}|\alpha|^2}.
\end{equation}
\end{example}
\begin{example}[Homodyne tomography in Gaussian noise] \cite{revt}
As for quantum efficiency, Gaussian noise can be
unbiased for mean thermal photon number $\bar n\le s-\frac 12$. One
has the new dual:
\begin{equation}
D(\alpha)\to\map{N}^{-1\dag}(D(\alpha))=D(\alpha)e^{\bar n|\alpha|^2}.
\end{equation}
\end{example}
\section{The case of homodyne tomography}
Before addressing the general problem of deriving a general
tomographic rule for unbounded operators, in this section we will
re-derive the known pattern function of homodyne tomography in order
to illustrate the general concepts introduced in the previous section.
\par The starting point is the observable frame $\{D(\alpha)\}_{\alpha\in\Cmplx}$, in terms of which
the decomposition (\ref{obsframe}) for Hilbert-Schmidt operators rewrites as follows
\begin{equation}
X=\int_\Cmplx
\frac{\d^2\alpha}{\pi}\Tr[D^\dag(\alpha)X]D(\alpha).\label{glauber}
\end{equation}
By changing to polar variables $\alpha = (-i/2)k e^{i\phi}$,  Eq. (\ref{glauber}) becomes 
\begin{equation}
X= \int^{\pi}_0\frac{d\phi}{\pi}\int^{+\infty}_{-\infty} 
\frac{d k\, |k|}{4}\,\Tr [ X\;  e^{ik X_{\phi}}]\, 
e^{-ik X_{\phi}}.\label{op}
\end{equation}
Eq. (\ref{op}) can be used only for Hilbert-Schmidt operators, for
which the trace under the integral in Eq. (\ref{op}) exists. In terms
of the quadrature spectral measure,  one has 
\begin{equation}
X= \int^{\pi}_0\frac{d\phi}{\pi}\int^{+\infty}_{-\infty} 
\frac{d k\, |k|}{4}\,\Tr [ X\;  e^{ik X_{\phi}}]\, e^{-ik X_{\phi}}
=\int^{\pi}_0\frac{d\phi}{\pi}\int^{+\infty}_{-\infty}\d E_\phi(x)
\Tr[X  W_{\phi,x}],
\end{equation}
where 
\begin{equation}
W_{\phi,x}=e^{-i\phi a^\dag a}D(x)W_{0,0}D^\dag(x)e^{i\phi a^\dag a},\qquad
W_{0,0}= -\frac 12\operatorname{P}\frac 1{X_0^2},\label{W}
\end{equation}
$\operatorname{P}$ denoting the Cauchy principal value. On the other hand, in the next section
we will show that for unbounded operators we also have the expansion
\begin{equation}
\begin{split}
X=&\int_0^{\pi}\frac{\d\phi}{\pi} \int_{-\infty}^\infty\d t
\Tr[G^\dag(t,\phi)X]F(t,\phi),\\
F(t,\phi)=&\frac{1}{\sqrt{2\pi}}e^{-\frac{1}{2}(t-i2X_\phi)^2},\quad
G(t,\phi)= \frac{\d}{\d t}t e^{\frac{t^2}{2}}  
\int_0^1\d\theta | i(1-\theta)te^{i\phi}\>\< -i\theta te^{i\phi}|.
\end{split}\label{Gexp}
\end{equation}
which in terms of the the quadrature spectral measure reads
\begin{equation}
X=\int_0^{\pi}\frac{\d\phi}{\pi}\int_{-\infty}^\infty 
\d E_\phi(x)\Tr[XW'_{\phi,x}],\label{frm1}
\end{equation}
where now
\begin{equation}
W'_{\phi,x}= \int_{-\infty}^\infty\d t
\frac{1}{\sqrt{2\pi}}e^{-\frac{1}{2}(t-i2x)^2} \frac{\d}{\d t}t
e^{\frac{t^2}{2}} \int_0^1\d\theta | i(1-\theta)te^{i\phi}\>\<
-i\theta te^{i\phi}|.\label{W1}
\end{equation}
\par Alternatively, as shown in Sec. (\ref{versus}) by means of the frame of normal-ordered moments,
one has the expansion
\begin{equation}
\begin{split}
&X=\sum_{n,m=0}^\infty  a^{\dag n}a^m \Tr [g_{n,m}^\dag X] \\ 
 &=  \int_{0}^\pi \frac{\d \phi }{\pi }\int_{-\infty }^{+ \infty }\d
E_\phi (x)\, \Tr \left [X \sum_{n,m=0 }^\infty g^\dag _{n,m} 
{ n+m  \choose n}^{-1} \frac{H_{n+m}(\sqrt 2\, x)}{\sqrt {2^{n+m}}}
\right ]\;\label{frm2}
\end{split}
\end{equation}
where 
\begin{eqnarray}
g_{n,m}=\sum_{j=0}^{\min{(n,m)}}
\frac{(-1)^{j}}{j!\sqrt {(n-j)!(m-j)!}}|n-j \> \<  m-j | 
\;.
\end{eqnarray}
The above expansions in Eqs. (\ref{frm1}) and (\ref{frm2}) are just
examples of alternate expansions which are equivalent for the
estimation of the expectation values of (even unbounded) observables,
but can be very different as regards the statistical noise affecting
such estimation.  As a matter of fact, the problem of classifying all
possible expansions has been never solved, and, hopefully, the results
of the present paper may suggest a unifying approach to the solution
of such a difficult problem. As we will see in the next section, the
existence of many alternate expansions is due to the symmetry of the
quorum of quadrature operators, and the resulting properties of null
estimator functions.
\section{Calculus with null functions}
We first notice that to a null estimator function $n_\xi\simeq 0$  it corresponds a null expansion over
the quorum, namely
\begin{equation}
\int_\Klm\d\mu(\xi)\int_\mathcal{X}\d E_\xi(x)n_\xi(x)=0\Longleftrightarrow
\int_\Klm\d\mu(\xi) n_\xi(X_\xi)=0.
\end{equation}
Let us recall the ordering relation \cite{wun3}
\begin{eqnarray}
\: a^{\dag k} a^l \: _s 
=\sum _{j=0}^{\{k,l\}} \frac{k ! l!}{j! (k-j)!
  (l-j)!}\left ( \frac {s-r}{2}\right )^j \: a^{\dag k-j}a^{l-j} \: _r\;,\label{arb}
\end{eqnarray}
where $\{k,l\}:=\min (k,l)$, and 
$s=-1,0,\hbox{ and }1$ correspond to normal, symmetrical, 
and anti-normal ordering, respectively. We will also write the symmetrical
ordering as $S\{ a^{\dag k} a^l\}:=\: a^{\dag k} a^l \: _0 $, and the
normal ordering as $\: a^{\dag k} a^l \: := \: a^{\dag k} a^l \: _{-1}
$. 

Then we have:  
\begin{lemma}[Main equivalence relation] \cite{null} 
The following equivalence relation holds
\begin{equation}
x^ke^{\pm i(k+2n+2)\phi}\simeq 0,\qquad \forall k,n\geq 0.\label{mainid}
\end{equation}
\end{lemma}
\Proof Since $X_\phi ^k = \frac{1}{2^k}\sum _{l=0}^k {k \choose l} S\{a^{\dag
  l} a^{k-l}\}\, e^{i\phi (2l-k)}$,  one has 
\begin{equation}
\int_0^{\pi}\!\!\frac{d\phi}{\pi} e^{\pm i(k+2+2n)\phi}
X_\phi^k=0,\qquad\forall k,n\geq 0, \label{mainidop}
\end{equation}
which is equivalent to (\ref{mainid}). 

\par Stated differently:
\begin{lemma} 
The following equivalence relation holds
\begin{equation}
H_k(\sqrt 2 \,x)e^{\pm i(k+2n+2)\phi}\simeq 0,\qquad \forall k,n\geq
0, \label{mainid2}
\end{equation}
where $H_k(x)$ denote the $k$-th Hermite polynomial.
\end{lemma}
\Proof From the definition of Hermite polynomials one has 
\begin{equation}
\begin{split}
\frac{1}{\sqrt {2^n}}
&H_n (\sqrt 2 X_\phi ) =
\frac{1}{\sqrt {2^n}}\left. \frac {\partial ^n}{\partial t^n}\right | _{t=0}
e^{-t^2 +\sqrt 2 t (a^\dag e^{i\phi }+a e^{-i\phi })} \\&=  
\frac{1}{\sqrt {2^n}}\left. \frac {\partial ^n}{\partial t^n}\right | _{t=0}
e^{\sqrt 2 t a^\dag e^{i\phi }}e^{\sqrt 2 t a e^{-i\phi }} =  
\sum _{k=0}^n {n \choose k} a^{\dag k} a^{n-k} e^{i\phi (2k-n)}=  2
^n {\bf :}X^n_\phi {\bf :}\;.\label{defh}
\end{split}
\end{equation}
Then, it follows that 
\begin{equation}
\int_0^{\pi}\!\!\frac{d\phi}{\pi} e^{\pm i(k+2+2n)\phi} H_k({\sqrt 2} X_\phi )=
0,\qquad\forall k,n\geq 0, 
\end{equation} 
which is equivalent to (\ref{mainid2}). 

\par Moreover, we also have 
\begin{lemma}[Equivalence of truncated Hermite polynomials]\label{l:Her} 
The following equivalence relations hold:
\begin{equation}
e^{\pm in\phi}H_{2l+n}^{(l)}(\kappa x)\simeq e^{\pm in\phi}H_{2l+n}(\kappa x),
\end{equation}
where we introduced the {\em truncated} Hermite polynomial
\begin{equation}
H_n^{(l)}(z)=\sum_{m=0}^l \frac{(-)^m n! (2z)^{n-2m}}{m!(n-2m)!}. \qquad n\ge 2l.  
\end{equation}
\end{lemma}
\Proof The two identities are just the complex conjugated of each other. Therefore, it is sufficient
to prove the identity with the plus sign. The latter is a simple consequence of identity (\ref{mainid}), namely
$x^ke^{\pm i(k+2n+2)\phi}\simeq 0$. 
By using the un-truncated Hermite polynomial, we have
\begin{equation}
\begin{split}
e^{in\phi}H_{2l+n}(\kappa x)=&
e^{in\phi}\sum_{m=0}^{l+[[n/2]]}\frac{(-)^m (2l+n)! (2\kappa)^{2l+n-2m}}{m!(2l+n-2m)!}\\
\times &x^{2l+n-2m}e^{i[n+2(l-m)+2(m-l)]\phi},
\end{split}
\end{equation}
from which it follows that all terms with $m>l$ are equivalent to zero.

Finally, one can show the Poisson identities (whose proof can be found
in the Appendix)
\begin{lemma}[Poisson identities]
The following identities hold
\begin{equation}
\begin{split}
f(x^2)\delta_\pi(\phi)&\simeq 
\frac{1}{\pi}\left[\frac{f(x^{2}e^{2i\phi})}{1-e^{-2i\phi}}+\frac{f(x^{2}e^{-2i\phi})}{1-e^{2i\phi}}\right],\\
xf(x^2)\delta_\pi(\phi)&\simeq \frac{1}{\pi}\left[\frac{xe^{i\phi}f(x^{2}e^{2i\phi})}{1-e^{-2i\phi}}+\frac{xe^{-i\phi}f(x^{2}e^{-2i\phi})}{1-e^{2i\phi}}\right],
\end{split}
\end{equation}
In particular, we have the identity
\begin{equation}
\delta_\pi(\phi)\simeq \frac{1}{\pi}.\label{dicomb} 
\end{equation}
\end{lemma}

\section{The Kolmogorov construction}
In this Section we present the Kolmogorov construction, and its
relation with the fundamental identity of quantum tomography. 

In the following, by $\ltwo$ we denote the Hilbert space of square
summable sequences of complex numbers, and by $\Ltwo(\Klm)$ we denote 
the Hilbert space of square summable functions over the space
$\Klm$. For example, $\Klm=\Reals$, and $\Ltwo(\Klm)$ the Hilbert space
of square summable functions on the real axis, or $\Klm=S^1$, and 
 $\Ltwo(\Klm)$ is the Hardy space of square-summable complex functions
on the circle. In the following we will focus attention 
only to the case $\Klm\equiv\Reals$. Consider now a complete orthonormal
set of functions $[\upsilon_n(x)]$ for $\Ltwo(\Klm)$. The completeness of the
set corresponds to the following distribution identity 
\begin{equation}
\sum_n \upsilon_n(x)^*\upsilon_n(y)=\delta(x-y),
\end{equation}
where $\delta$ denotes the usual Dirac-delta. Consider now a
(infinite-dimensional) Hilbert space $\sH$ and denote by $[w_n]$ an
orthonormal basis for it. The following vector 
\begin{equation}
|\upsilon(x)\>=\sum_n\upsilon_n(x)|w_n\>,
\end{equation}
is Dirac-normalizable, in the sense that
\begin{equation}
\<\upsilon(y)|\upsilon(x)\>=\delta(x-y).
\end{equation}
Consider now another (infinite-dimensional) Hilbert space
$\sK\simeq\sH$. To the orthonormal basis $[\upsilon_n(x)]$ for
$\Ltwo(\Klm)$ and $[z_n]$ for $\sK$ we associate a map from the
observables $\set{O}_\Klm$ with spectrum $\Klm$ on  
$\sH$ to operators in  $\Bnd{\sH,\sH\otimes\sK}$ given by 
\begin{equation}
\upsilon(X)=\sum_n\upsilon_n(X)\otimes|z_n\>,
\end{equation}
where, as usual,  we define the operators $\upsilon_n(X)$ in terms of the spectral
resolution of $X$ 
as follows
\begin{equation}
\upsilon_n(X)=\int_\Klm\d E_X(x)\upsilon_n(x),
\end{equation}
where $\d E_X(x)$ denotes the spectral measure of $X$. 
\par For $X,Y\in\set{O}_\Klm$ formally we write
\begin{equation}
\upsilon(X)^\dag\upsilon(Y)=\sum_n \upsilon_n(X)^\dag \upsilon_n(Y),
\end{equation}
Consider the integral kernel $K(x,y)$, $x,y\in\Klm$ corresponding 
to the positive operator $K\in\Bnd{H}$
\begin{equation}
K(x,y)=\<\upsilon(x)|K|\upsilon(y)\>.
\end{equation}
For any two self-adjoint operators $X,Y$ on $\Ltwo(\Reals)$, the
expression $K(X,Y)$ is well defined in the following sense 
\begin{equation}
K(X,Y)\doteq \upsilon(X)^\dag (I\otimes\tilde{K}) \upsilon(Y),
\end{equation}
where $\tilde{K}\in\Bnd{K}$ is given by
\begin{equation}
\tilde{K}=\sum_{n,m} |z_n\>\<w_n|K|w_m\>\<z_m|,
\end{equation}
namely we can also write
\begin{equation}
K(X,Y)=\sum_{n,m}\upsilon_n(X)^\dag \<w_n|K|w_m\> \upsilon_m(Y) 
=\int_\Klm\d E_X(x) \int_\Klm\d E_X(y) K(x,y).\label{pro}
\end{equation}
This is also equivalent to say that for any expansion of
$K(x,y)$ in series of products of functions of single variable,  
$K(X,Y)$ is defined as the same expansion, ordered with the functions
of $X$ on the left and the functions of $Y$ on the right. For
commuting $X,Y$, then  $K(X,Y)$ simply represents the same analytic
expression of $K(x,y)$, now substituting the operators in place of the
variables. As an example, the identity operator $K=I$ corresponds to the
Dirac-delta kernel, and for commuting $X,Y\in\set{O}_\Klm$ we have  
\begin{equation}
\upsilon(X)^\dag\upsilon(Y)=\delta(X-Y).
\end{equation}
By replacing now $\sH\to\sH^{\otimes 2}$, even for non commuting $X$
and $Y$, one has 
\begin{equation}
(\upsilon(X)^\dag \otimes I )(I \otimes \upsilon(Y))=\delta(X\otimes I-I\otimes Y).
\end{equation}
Moreover, similarly to Eq. (\ref{pro}),  one has  
\begin{eqnarray}
K(X \otimes I ,I \otimes Y) &\doteq &
(\upsilon(X)^\dag \otimes I )
(I_\sH^{\otimes 2}\otimes  \tilde{K}) (I \otimes \upsilon(Y)) 
\nonumber \\&= &
\sum_{n,m}\upsilon_n(X)^\dag \otimes \upsilon_m(Y)   \<w_n|K|w_m\> 
\nonumber \\&= & 
\int_\Klm\d E_X(x) \otimes \int_\Klm\d E_X(y) K(x,y).\label{pro2}
\end{eqnarray}
The fundamental identities of quantum tomography are obtained as an
expansion of the swap operator $E$ over the {\em quorum}, since for
any state $\rho $ and observable $A$ one has $\Tr[\rho A]=\Tr[(\rho
\otimes A) E]$, where $E|\psi \> \otimes |\phi \> =|\phi
\> \otimes |\psi \> $. 
\subsection{Homodyne tomography}
From Eq. (\ref{op}), it is clear the swap operator can be written as 
\begin{equation}
E= \int^{\pi}_0\frac{d\phi}{\pi}\int^{+\infty}_{-\infty} 
\frac{d k\, |k|}{4}\,e^{-ik X_{\phi}} \otimes  
e^{ik X_{\phi}}.\label{op2}
\end{equation}
Then, the usual homodyne tomographic formula can be obtained by the
Kolmogorov construction in writing the swap
operator as follows
\begin{equation}
E=\int_0^\pi\frac{\d\phi}{\pi} (\upsilon(X_\phi)^\dag \otimes I)
(I_\sH^{\otimes 2}\otimes \tilde {K}) 
I \otimes \upsilon(X_\phi),
\end{equation}
corresponding to the positive kernel
\begin{equation}
K(x,x')=\frac{\pi}{2}\< x||Y||x'\>=\int_\Reals\frac{\d k}{4}|k|e^{ik(x'-x)}
=-\frac{\operatorname{P}}{2} \frac{1}{(x-x')^2},  
\end{equation}
where $Y$ is the quadrature conjugated to  $X$, with $[X,Y]=\frac i
2$. 
The kernel is clearly positive, since one has
\begin{equation}
\sum_{i,j}K(x_i,x_j)\xi_i^*\xi_j=
\int_\Reals\frac{\d k}{4}|k|\left|\sum_je^{ix_j}
\xi_j\right|^2
\end{equation}
The tomographic formula consists in the following identity
\begin{equation}
E=\int_0^\pi\frac{\d\phi}{\pi} \, K(X_\phi\otimes I,I\otimes X_\phi) 
\end{equation}
Using Eq. (\ref{pro2}), one can also write 
\begin{eqnarray}
E= \int_0^\pi\frac{\d\phi}{\pi} \sum _n \upsilon _n (X_\phi)^\dag \otimes
u_n (X_\phi)\;,
\end{eqnarray}
with 
\begin{eqnarray}
u_n(X_\phi)=\sum _m \upsilon _m (X_\phi) \< n | |Y| |m \>
\;. 
\end{eqnarray}
The existence of null estimator functions can be taken into account by
considering any operator $N_{n,\phi}$ such that 
\begin{eqnarray}
\int_0^\pi\frac{\d\phi}{\pi} \upsilon _n (X_\phi)^\dag \otimes N_{n,\phi}=0
\;,
\end{eqnarray}
and any estimation rule can be obtained by the swap operator
\begin{eqnarray}
E= \int_0^\pi\frac{\d\phi}{\pi} \sum _n \upsilon _n (X_\phi)^\dag \otimes
D_{n,\phi}\;,
\end{eqnarray}
with $D_{n,\phi}=u_n(X_\phi)+ N_{n,\phi}$, as follows
\begin{eqnarray}
\Tr[\rho X] =\int_0^\pi\frac{\d\phi}{\pi} \sum _n \Tr [\rho \upsilon
_n (X_\phi)^\dag ]\Tr [D_{n,\phi} X]\;.
\label{rull}
\end{eqnarray}
\subsection{Spin tomography}
For spin tomography the swap operator writes as follows
\begin{equation}
E=\frac{2J+1}{2\pi}\int\frac{\d\vec n}{4\pi}\int_0^{2\pi}\d\psi
\sin^2 \tfrac{\psi}{2} e^{i(\vec J_1-\vec J_2)\cdot\vec n\psi}=
\int\frac{\d\vec n}{4\pi} K(\vec J_1\cdot\vec n \otimes I, I\otimes 
\vec J_2\cdot\vec n),    
\end{equation}
and we immediately see that the kernel can be written as follows
\begin{equation}
K(r,s)=\tfrac{1}{2}\left(J+\tfrac{1}{2}\right)\< r+J|2-e_+-e_-|s+J\>,
\end{equation}
where $r,s=-J,-J+1\ldots J$, and 
$\{|n\> \}$ denote any orthonormal basis of the infinite dimensional
Hilbert space $\sH$. Such a basis can be conveniently regarded
as the Hardy space of functions on the unit circle, with 
$\<n|z\>=z^n$,  $|z|=1$, and 
\begin{equation}
\oint \frac{\d z}{2\pi i z} |z\>\< z|\equiv
\int_0^{2\pi}\frac{\d\psi}{2\pi} |e^{i\psi}\>\< e^{i\psi}|,
\end{equation}
$e_-$ denoting the shift operator
$e_-|n\>=|n-1\>$, and $e_+=e_-^\dag$.
By introducing the vectors $|\upsilon(m)\>\doteq |m+J\>$, we can write
\begin{equation}
E=\int\frac{\d\vec n}{4\pi} (\upsilon(\vec J_1\cdot\vec n)^\dag \otimes
I)(I_{\sH^{\otimes2}}
\otimes K)(I\otimes \upsilon(\vec J_2\cdot\vec n)),
\end{equation}
where the operator $K\in\Bnd{K}$ is given by
\begin{equation}
K=\left(J+\tfrac{1}{2}\right)(1-C), 
\end{equation}
where $C=\frac{1}{2}(e_++e_-)$ is the \Index{cosine operator}.

\subsection{Alternate expansions}
The general form of the swap operator is
\begin{equation}
E=\sum_\nu (\upsilon(X_\nu )^\dag \otimes I)
(I_{\sH^{\otimes 2}}\otimes K)(I \otimes \upsilon(X_\nu)).
\end{equation}
Introducing an invertible operator $L\in\Bnd{K}$, we can write
\begin{eqnarray}
E&=&\sum_\nu (\upsilon(X_\nu )^\dag \otimes I)
(I_{\sH^{\otimes 2}}\otimes K^{\frac{1}{2}}L^{-1}LK^{\frac{1}{2}})(I \otimes \upsilon(X_\nu)),
\nonumber \\&= & 
\sum_\nu \sum_{n,m,l}
\upsilon_n(X_\nu)^\dag\otimes\upsilon_m(X_\nu)
\<n|K^{\frac{1}{2}}L^{-1}|z(l)\>\<z(l)|LK^{\frac{1}{2}}|m\>,
\;
\end{eqnarray}
where $\{|z(l)\>\}$ is any orthonormal basis for $\sK$. Therefore,
we have all the alternate expansions on the quorum
\begin{equation}
Z=\sum_\nu \sum_l \Tr[L_l(X_\nu)^\dag Z]M_l(X_\nu),\label{Zexp}
\end{equation}
where
\begin{equation}
\begin{split}
M_l(x)&=\sum_m\upsilon_m(x)\<z(l)|LK^{\frac{1}{2}}|m\>
=\<z(l)|LK^{\frac{1}{2}}|\upsilon(x)\>,\\
L_l(x)^*&=\sum_m\upsilon_m^*(x)\<m|K^{\frac{1}{2}}L^{-1}|z(l)\>=
\<\upsilon(x)|K^{\frac{1}{2}}L^{-1}|z(l)\>.
\end{split}
\end{equation}

\section{Expansion of unbounded operators over the quadratures}
As already noticed, the swap operator in Eq. (\ref{op2}) provides the
estimation rule just for trace-class operators. However, it is known
since Richter \cite{Rich} the following formula
\begin{eqnarray}
a^{\dag n} a^m=
{n+m \choose n}^{-1}\,\int_{0}^\pi \frac {\d\phi }{\pi }
\,\frac {1}{\sqrt {2^{n+m}}} \,H_{n+m}(\sqrt 2 X_\phi )
\,e^{i\phi (m-n)}
\;.\label{ric}
\end{eqnarray}
Eq. (\ref{ric}) was originally derived by using nontrivial identities 
involving trilinear products of Hermite polynomials. 

Here, we provide a much simpler derivation as follows. Using 
the definition of Hermite polynomials in Eq. (\ref{defh}),  one has 
\begin{equation}
\begin{split}
a^{\dag n} a^m&={n+m \choose n}^{-1}\,\sum _{k=0}^{n+m}{n+m\choose
  k}a^{\dag k} a^{n+m-k} \,\delta _{k,n} \\&=  
{n+m \choose n}^{-1}\,\int_{0}^\pi \frac {\d\phi }{\pi }
\sum _{k=0}^{n+m}{n+m\choose
  k}a^{\dag k} a^{n+m-k} \,e^{i\phi (2k-n-m)}e^{i\phi (m-n)}
\\&=  
{n+m \choose n}^{-1}\,\int_{0}^\pi \frac {\d\phi }{\pi }
\,{\bf :}X^{n+m}_\phi {\bf :}
\,e^{i\phi (m-n)}
\\&= 
{n+m \choose n}^{-1}\,\int_{0}^\pi \frac {\d\phi }{\pi }
\,\frac {1}{\sqrt {2^{n+m}}} \,H_{n+m}(\sqrt 2 X_\phi )
\,e^{i\phi (m-n)}
\;\label{otto}
\end{split}
\end{equation}
Similarly, for the symmetrical ordering, one derives the identity
\begin{eqnarray}
S\{ a^{\dag n} a^m\} &=& {n+m \choose m }^{-1}\sum _{l=0}^{n+m} {n+m
  \choose l} S\{ a^{\dag n+m-l} a^l\} \delta _{lm} 
\nonumber \\&= &
 {n+m \choose m }^{-1}\int _{0} ^\pi \frac {d\phi }{\pi} e^{i\phi
   (m-n)}
\sum _{l=0}^{n+m} {n+m
  \choose l} S\{ (a^{\dag }e^{i\phi })^ {n+m-l} (ae^{-i\phi
})^l\}\nonumber \\&= &
 {n+m \choose m }^{-1}\int _{0} ^\pi \frac {d\phi }{\pi} 
e^{i\phi
   (m-n)} \, 2^{n+m}\, X_\phi ^{n+k} 
\;.\label{simm}
\end{eqnarray}
For arbitrary ordering, using Eq. (\ref{arb}), one obtains 
\begin{eqnarray}
:a^{\dag k} a^l: _s 
&=&\sum _{j=0}^{\{k,l\}} \frac{k ! l!}{j! (k-j)!
  (l-j)!}\left ( \frac {s}{2}\right )^j S\{a^{\dag k-j}a^{l-j} \}
\nonumber \\&= &
\int _{0} ^\pi \frac {d\phi }{\pi}  e^{i\phi (l-k)}
\sum _{j=0}^{\{k,l\}} \frac{k ! l!}{j! (k+l-2j)!} \left (\sqrt {\frac
    {s}{2}}\right )^{k+l}
\left( 2\sqrt {\frac 2
    s} X_\phi \right )^{k+l-2j}\nonumber \\&= &
{ k+l \choose k }^{-1} 
\int _{0} ^\pi \frac {d\phi }{\pi}  e^{i\phi (l-k)}
\left (\sqrt {\frac
    {s}{2}}\right )^{k+l}
H_{k+l}^{(k,l)}\left (\sqrt {\frac 2s} X_\phi \right )
\;.
\end{eqnarray} 
Using Lemma 3, one has the equivalent identity
\begin{eqnarray}
:a^{\dag k} a^l: _s =
{ k+l \choose k }^{-1} 
\int _{0} ^\pi \frac {d\phi }{\pi}  e^{i\phi (l-k)}
\left (\sqrt {\frac
    {s}{2}}\right )^{k+l}
H_{k+l}\left(\sqrt {\frac 2s} X_\phi \right)
\;.
\end{eqnarray}  
In a similar way, one can derive the useful relation
\begin{eqnarray}
(\mu a + \nu a^\dag )^n =\int _0 ^\pi \frac{d\phi }{\pi} (2 X_\phi )^n  \frac{(\nu
  e^{-i\phi})^{n+1}- (\mu e^{i\phi})^{n+1}}{\nu e^{-i\phi }-\mu e^{i\phi}}
\;, 
\end{eqnarray}
whence
\begin{eqnarray}
X_\varphi ^n =\int _0 ^\pi \frac{d\phi }{\pi} X_\phi ^n 
\frac{\sin [(\phi -\varphi)(n+1)]}{\sin (\phi -\varphi)}
\;.
\end{eqnarray}
Using Eq. (\ref{simm}), for the displacement operator one obtains
\begin{eqnarray}
D(\alpha )&=&\sum _{n=0}^\infty \frac {1}{n !}(\alpha a^\dag -\alpha ^*
a)^n = \sum _{n=0}^\infty \sum _{k=0}^n {n\choose k} \alpha
^{n-k}(-\alpha ^*)^k S\{a^{\dag n-k} a^k\} 
\nonumber \\&= &
\int _{0} ^\pi \frac {d\phi }{\pi} 
\sum _{k=0}^\infty \sum _{n=0}^\infty  \frac {(2 \alpha e^{-i\phi
  }X_\phi)^n (-2 \alpha ^* e^{i\phi} X_\phi)^k}{(n+k)!}
\;.\label{estr}
\end{eqnarray}
The last equation can be summed using the identity
\begin{equation}
\begin{split}
&\sum_{n,m=0}^\infty \frac{z^n(-z^*)^m}{(n+m)!}=\sum_{s=0}^\infty \frac{z^s}{s!}\sum_{d=0}^s\left(
-\frac{z^*}{z}\right)^d=\sum_{s=0}^\infty\frac{z^s}{s!}
\frac{1-\left(-\frac{z^*}{z}\right)^{s+1}}{1+\frac{z^*}{z}}\\
=&\sum_{s=0}^\infty\frac{1}{s!}\frac{z^{s+1}-(-z^*)^{s+1}}{z+z^*} =\frac{ze^z+z^*e^{-z^*}}{z+z^*}  
\end{split}
\end{equation}
which gives the estimation rule
\begin{equation}
f_\phi(X_\phi|D(\alpha))=\frac{e^{-i\phi}\alpha e^{2X_\phi e^{-i\phi}\alpha}
+e^{i\phi}\alpha^*e^{-2X_\phi
  e^{i\phi}\alpha^*}}{e^{-i\phi}\alpha+e^{i\phi}\alpha^*}  \;.
\label{estr2}
\end{equation}
Identity (\ref{estr2}) should be compared with the equivalent estimator given in Ref. \cite{tokio}.
The identity in Eq. (\ref{estr}) can be also derived by explicitly using the properties of null
estimator functions, as shown in the Appendix.

\par All estimation rules $f_\phi(X_\phi | X)$ given in the present section do not
correspond to an expectation of $X$ as in Eq. (\ref{rull}). 
However, we can suitably recover an expectation rule---which is
generally not unique---for any observable.   
Consider, for example, Eq. (\ref{estr}). Using 
the following identity \cite{gradshtein}
\begin{equation}
\int_0^1\d\theta\theta^n\,(1-\theta)^m=B(n+1,m+1)=\frac{n!m!}{(n+m+1)!}, 
\end{equation}
along with
\begin{equation}
(m+n+1)=\left.\frac{\d}{\d t} t^{m+n+1}\right|_{t=1},
\end{equation}
one obtains the integral form for the inverse binomial coefficient 
\begin{equation}
{ m+n \choose m}^{-1} 
=\left.\frac{\d}{\d
    t}\right|_{t=1}t\,\int_0^1\d\theta(t\theta)^n\,(t(1-\theta))^m .
\label{invbin}
\end{equation}
Then, the estimator (\ref{estr}) becomes
\begin{equation}
f_\phi(X_\phi|D(\alpha))=\left.\frac{\d}{\d t}\right|_{t=1}t\, \int_0^1\d\theta
\exp(-2 X_\phi e^{i\phi}\alpha^*t\theta)
\exp(2 X_\phi e^{-i\phi}\alpha t(1-\theta)),
\end{equation}
and for its spectral kernel one has
\begin{equation}
f_\phi(x|D(\alpha))=\frac{\d}{\d x}x\, \int_0^1\d\theta\exp(-2 x e^{i\phi}\alpha^*\theta)
\exp(2 x e^{-i\phi}\alpha (1-\theta)).
\end{equation}
We can rewrite the estimator in form of expectation 
\begin{equation}
f_\phi(x|D(\alpha))=\frac{\d}{\d x}x\, \int_0^1\d\theta
\< 2 x e^{i\phi}(1-\theta)|e^{\alpha a^\dag}e^{-\alpha^* a}|2 x e^{i\phi}\theta\>,\label{form}
\end{equation}
where the vectors are coherent states. 
Eq. (\ref{form}) corresponds to the functional form
\begin{equation}
f_\phi(x|D(\alpha))=\Tr[W_{\phi,x}\:D(\alpha)\:], 
\end{equation}
where
\begin{equation}
W_{\phi,x}=\frac{\d}{\d x}x\, \int_0^1\d\theta|2 x e^{i\phi}\theta\>\< 2 x e^{i\phi}(1-\theta)|.
\end{equation}
In order to give an expectation rule corresponding to
Eq. (\ref{otto}), we can proceed as follows. 
From the integral representation \cite{gradshtein} 
\begin{equation}
H_n(x) = (-2i)^n\int_{-\infty}^\infty \frac{\d t}{\sqrt{\pi}}
e^{-(t-ix)^2}\,t^n\;,
\end{equation}
using identity (\ref{invbin}), Eq. (\ref{otto}) for normal ordering rewrites as follows
\begin{equation}
\begin{split}
a^{\dag m} a^n = &\int_0^{\pi}\frac{\d\phi}{\pi}
\int_{-\infty}^\infty\frac{\d t}{\sqrt{2\pi}}
e^{-\frac{1}{2}(t-i2X_\phi)^2}
\frac{\d}{\d t} t \int_0^1\d\theta (-i\theta te^{i\phi})^n
[-i(1-\theta)te^{-i\phi}]^m\\ =&
\int_0^{\pi}\frac{\d\phi}{\pi} \int_{-\infty}^\infty\frac{\d t}{\sqrt{2\pi}}
e^{-\frac{1}{2}(t-i2X_\phi)^2}\\ \times &
\frac{\d}{\d t}t \int_0^1\d\theta \Tr [a^{\dag m} a^n  | -i\theta
te^{i\phi}\>\< i(1-\theta)te^{i\phi}|]e^{\frac{t^2}{2}},\label{boring}
\end{split}
\end{equation}
where we have used matrix elements on coherent states
\begin{equation} 
\<\beta| a^{\dag m} a^n |\alpha\> =\alpha ^n{\beta ^*  m}
e^{-\frac{1}{2}(|\alpha|^2+|\beta|^2-2\alpha\beta^*)}.
\end{equation}
Eq. (\ref{boring}) is equivalent to the following expansion 
for operators $X$ admitting normal ordered
form
\begin{equation}
\begin{split}
X=&\int_0^{\pi}\frac{\d\phi}{\pi} \int_{-\infty}^\infty\d t
\Tr[G^\dag(t,\phi)X]F(t,\phi),\\
F(t,\phi)=&\frac{1}{\sqrt{2\pi}}e^{-\frac{1}{2}(t-i2X_\phi)^2},\quad
G(t,\phi)= \frac{\d}{\d t} t e^{\frac{t^2}{2}}  
\int_0^1\d\theta | i(1-\theta)te^{i\phi}\>\< -i\theta te^{i\phi}|.
\end{split}
\end{equation}
\section{Canonical dual for homodyne tomography}  
The frame theory approach to quantum homodyne tomography gives further insight
to the structure of the {\em quorum} of quadrature observables. Given
a set of vectors ${|v_n \> }$ in a Hilbert space, if the positive operator $F=\sum
_{n}|v_n \> \< v_n |$ is invertible, then the scalar product
between two arbitrary vectors can be written as 
\begin{eqnarray}
\< \psi | \eta \> = \sum _n \< \psi | u_n \>
\< v_n  | \eta \>  \;,
\end{eqnarray}
where the set of vectors $\{u_n \equiv F^{-1}|v_n \> \}$ is called
``canonical dual'' of the set $\{ |v_n \> \}$, and $F$ is denoted
as ``frame operator''. In other words, the set $\{|v_n \> \}$,
along with its dual $\{|u_n \> \}$, is a spanning set for the
Hilbert space, and provide a generalized resolution of the identity. 
In this section we show that the set of
(generalized) projectors $|x \> _\phi  {}_\phi \< x| $ over
the quadratures $X_\phi $ give a frame when varying $\phi $, and the
expansion for trace-class operators in Eq. (\ref{op}) corresponds to
using the canonical dual for the estimation rule. 

In the following, we will make extensive use of the isomorphism
between the Hilbert space of the Hilbert-Schmidt
operators $A,B$ on $\sH$, with scalar product $\< A,B\>
=\hbox{Tr}[A^\dag B]$, and the Hilbert space of bipartite vectors
$|A{\>\!\>},|B{\>\!\>}\in {\sH}\otimes{\sH}$, with
${\<\!\<} A|B{\>\!\>}\equiv\<
A,B\> $, and
\begin{eqnarray}
|A{\>\!\>}=\sum_{n}
\sum_{m} 
A_{nm}|n\> \otimes|m\> \;,\label{iso} 
\end{eqnarray}
where $|n\>$ and $|m\>$ are fixed orthonormal bases for
$\sH$, and $A_{nm}=\< n|A|m \>
$. Notice the identities
\begin{eqnarray}
A\otimes B\dk{C}=\dk{ACB^\tau}\,,\qquad A\otimes B^\dag \dk{C}=\dk{ACB^*}\,,
\label{ids}
\end{eqnarray}
where $\tau$ and $*$ denote transposition and complex conjugation 
with respect to the fixed bases in Eqs. (\ref{iso}). 

By taking $\< n|x \> _0$ as real,  
in the $|  \ \kk $ notation the spanning set 
$|x \> _\phi {}_\phi \< x|$ corresponds to the vectors on
$\sH _a \otimes \sH _b$ of modes $a$ and $b$
\begin{equation}
| (|x \> _\phi {}_\phi \<
x|) \kk = e^{i\phi (a^\dag a -b^\dag b)}| x \> _0 |x \> _0 
\;.
\end{equation}
Notice the identities  
\begin{equation}
\bb D(z)| x \> _0 |x \> _0 =
\exp (2i x \hbox{Im} z) \,\delta
(\hbox{Re} z) \;,
\end{equation}
and
\begin{equation}
e^{i \phi (a^\dag a -b^\dag b)}|D(z) \kk =|D(z e^{i\phi}) \kk \;.
\end{equation}
The frame operator can be evaluated as follows 
\begin{equation}
\begin{split}
F&=\int _0^\pi \frac{\d\phi }{\pi }\int _{-\infty}^{\infty} \d x\,
e^{i\phi (a^\dag a -b^\dag b)}| x \> _0 |x \> _0 {}_0\<
x |{}_0 \< x | e^{-i\phi (a^\dag a -b^\dag b)} \\ &=
\int _0^\pi \frac{\d\phi }{\pi }\int _{-\infty}^{\infty} \d x\, \int
\frac{\d^2 z }{\pi }\int \frac{\d^2 w}{ \pi} |D(z) \kk \bb
D(ze^{-i\phi})| x \> _0 |x \> _0 
{}_0\< x |{}_0 \< x | D(we^{-i\phi }) \kk \bb D(w)|
\\ & = \int _0^\pi \frac{\d\phi }{\pi }\int
_{-\infty}^{\infty} \d x\, \int \frac{\d^2 z }{\pi }\int \frac{\d^2 w}{
\pi} |D(ze^{i\phi}) \kk \bb D(we^{i\phi }) | e^{2ix
\hbox{\scriptsize Im}z - 2ix
\hbox{\scriptsize Im}w } \delta
(\hbox{Re}z) \delta (\hbox{Re}w) \\ &= \frac 1 \pi \int
_0^\pi \frac{\d\phi }{\pi }\int _{-\infty}^{\infty} dt\, |D(it e^{i\phi
}) \kk \bb D(ite^{i\phi })| = \frac {1}{\pi} \int\frac{\d^2 z}{\pi}\frac {1}{|z|}
|D(z) \kk 
\bb D(z)|\\&= \frac{1}{\pi|a-b^\dag|} \;, \label{4} 
\end{split}
\end{equation}
where we used the eigenvalue equation $(a- b^\dag )|D(z) \kk =z |D(z)
\kk $. 
The inverse of $F$ is simply given by 
\begin{equation}
F^{-1}=\pi|a-b^\dag|=\int \d^2 z \,|z|\,
|D(z) \kk \bb D(z)|\;.\label{5}
\end{equation}
The canonical dual is obtained as follows
\begin{equation}
\begin{split}
&F^{-1} | (|x \> _\phi {}_\phi \< x|) \kk =
\int \d^2 z \,|z| |D(z) \kk \bb D(z)| x \> _\phi |x \>
_\phi \\&=\int \d^2 z \,|z|\, |D(ze^{i\phi}) \kk
e^{2i x \hbox{\scriptsize Im}z }\delta (\hbox{Re}z) =
\int_{-\infty }^\infty \d k\,|k|\,|D(ike^{i\phi}) \kk e^{2ikx}
\\ & =\frac 14 \int_{-\infty }^\infty
\d k\,|k|\,|e^{ik(X_\phi -x)} \kk \;.\label{6}
\end{split}
\end{equation}
Hence, it follows that the usual Kernel operator corresponds to 
the canonical dual. 

\subsection{Alternate dual frames}
The dual of the quadrature projectors is not unique. However, the formula of Li \cite{Li} for
characterize all possible alternate duals for bounded frames and discrete indexes cannot provide any
new dual set.  By denoting the frame as $\{|\Xi(x,\phi)\>\!\>\}$ with $\Xi(x,\phi) = \delta (X_\phi
-x)$, such a formula can be formally written in the form
\begin{equation}
|\Theta(x,\phi)\>
\!\>=F^{-1}|\Xi(x,\phi)\>\!\>+|f(x,\phi)\>\!\>-
\int_0^\pi \frac{\d\phi'}{\pi} \int_{-\infty}^\infty d x'
\<\!\<\Xi(x',\phi')|F^{-1}|\Xi(x,\phi)\>\!\>
|f(x',\phi')\!\>\,,
\label{allduals}
\end{equation}
where $\{F^{-1}|\Xi(x,\phi)\>\!\>\}$ is the canonical dual, and
$\{f(x,\phi)\}$ is an arbitrary Bessel set, namely 
\begin{equation}
\int_{-\infty}^\infty \d x\int_0^\pi \frac{\d\phi}{\pi} |f(x,\phi)|^2\le \infty.
\end{equation}
The scalar product that appears in the integral of
Eq. (\ref{allduals}) writes 
\begin{eqnarray}
&&\<\!\<\Xi(x',\phi')|F^{-1}|\Xi(x,\phi)\>\!\>
\\& &  =
\int d^2 z \,|z| {}_0 \< x' | {}_0 \< x'|
|D(z) \kk \bb D(ze^{i(\psi -\phi)})| x \> _0 |x \> _0 
\nonumber \\& & =\int d^2 z \,|z|\,
e^{-i \hbox{\scriptsize Im}z (\hbox{\scriptsize Re}z+2x')}\,\delta
(\hbox{Re}z) 
e^{i \hbox{\scriptsize Im}(ze^{i(\psi -\phi)})(\hbox{\scriptsize
    Re}(ze^{i(\psi -\phi)}+2x)}\,\delta (\hbox{Re}(ze^{i(\psi -\phi)})) 
\nonumber \\& &=
\int _{-\infty}^{\infty }dk\,|k|\,
e^{2i k[x\cos (\psi -\phi )-x']}
\,\delta(k\sin (\psi -\phi))
\nonumber \\& &=
\int _{-\infty}^{\infty }dk\,
e^{2i k[x\cos (\psi -\phi )-x']}
\,\delta_\pi (\psi -\phi)
\nonumber \\& & =
\pi \,\delta(x-x')\,\delta_\pi (\psi -\phi)
\;.\label{wr}
\end{eqnarray}
This bi-orthogonality relation implies that the formula
(\ref{allduals}) cannot reveal any new dual set. 
\section{Frames of normal-ordered moments}
In this Section, by simply applying the frame theory, we recover some results of Refs.
\cite{Wun1,Lee}, where the set of normally ordered moments $\{a^{\dag k}a^l\}$ is shown to be
complete, and related to a biorthogonal set given on the basis of Fock states.
\par From the set $\{a^{\dag k}a^l\}$ we immediately write the frame
operator 
\begin{equation}
\tilde F=\sum_{k,l=0}^\infty  \dk {a^{\dag k}a^l} \db{a^{\dag k}a^l}
\;.\label{fnorm}
\end{equation}
On the Fock basis one has
\begin{equation}
\begin{split}
\tilde F&=  
\sum_{k,l,n,j=0}^\infty  \frac{\sqrt{(k+n)!(l+n)!(k+j)!(l+j)!}}{n! j!} 
|k+n \> |l+n \> \< k+j |\< l+j | \\ &=\sum_{n=0}^\infty \frac{a^nb^n }{n!} 
(\sum_{k,l=0}^\infty  k! l!|k \> \< k|\otimes |l \>
\< l |) \sum_{j=0}^\infty \frac{a^{\dag j}b^{\dag j}}{j!} \\&= 
e^{ab} (a^\dag a !\otimes b^\dag b !) e^{a^{\dag }b^{\dag}}.
\end{split}
\end{equation}
The inverse of $\tilde F$ simply writes 
\begin{equation}
\tilde F^{-1}=e^{-a^\dag b^\dag} 
\left(\frac {1}{a^\dag a !}\otimes \frac{1}{b^\dag b!} \right) e^{-ab}
\;.
\end{equation}
The dual set $g_{k,l}$ can be obtained as follows 
\begin{equation}
\begin{split}
\dk{g_{k,l}}&\equiv  \tilde F^{-1}\dk{a^{\dag k}a^l }
=\sum_{n,m,t,j=0}^{\infty }
|n \> |m \> \frac{(-1)^{t+j}}{t!j!} \< n|\< m|
a^t b^t \frac{1}{a^\dag a!}\otimes \frac{1}{b^\dag b!} a^{\dag j}b^{\dag
  j}\dk{a^{\dag k}a^l }
\\& = \sum_{n,m,t,j,s=0}^{\infty }
|n \> |m \> \frac{(-1)^{t+j}}{t!j!} \< n+t|k+j+s
\> \< m+t| l+j+s \>
\frac{\sqrt{(k+j+s)!(l+j+s)!}}{s!\sqrt{n!m! (n+t)!(m+t)!}}
\\& = \sum_{n,m,t,j,s=0}^{\infty }
|n \> |m \> \frac{(-1)^{t+j}}{t!j!} \delta
_{n+t,k+j+s}\delta _{m+t,l+j+s} 
\frac{\sqrt{(k+j+s)!(l+j+s)!}}{s!\sqrt{n!m! (n+t)!(m+t)!}}
\\&= \sum_{n,m,t=0}^{\infty }
\frac{1}{\sqrt {n!m!}}|n \> |m \> \frac{(-1)^{t}}{t!} 
\delta_{m,l+n-k} \frac{1}{(n+t-k)!}  
\sum _{j=0}^{n+t-k}\frac{(-1)^{j}}{j!}\frac{(n+t-k)!}{(n+t-k-j)!} 
\\& =\sum_{n,m,t=0}^{\infty }
\frac{1}{\sqrt {n!m!}}|n \> |m \> \frac{(-1)^{t}}{t!} 
\delta _{n,k-t}\delta_{m,l-t} \frac{1}{(n+t-k)!}  
\\&= \sum_{t=0}^{\min{(k,l)}}
\frac{(-1)^{t}}{t!\sqrt {(k-t)!(l-t)!}}|k-t \> | l-t \> 
\;.
\end{split}
\end{equation}
The dual set is then given by 
\begin{equation}
g_{k,l}=\sum_{t=0}^{\min{(k,l)}}
\frac{(-1)^{t}}{t!\sqrt {(k-t)!(l-t)!}}|k-t \> \<  l-t | =\sum_{t=0}^\infty
\frac{(-1)^t}{t!} a^\dag{}^{k-t}|0\> \<  0|a^{l-t}.
\end{equation}
The dual set is unique, and in fact one has the biorthogonal relation
\begin{equation}
\hbox{Tr}[g^{\dag} _{k',l'}a^{\dag k}a^l]=\delta _{k,k'}\delta _{l,l'}.
\;
\end{equation}
\section{Frame of moments versus quadrature distribution}\label{versus}
The frame of moments allows to recover the estimation rule for
unbounded operators as an expectation rule with a dual operator of the
quadrature projectors $|x \> _\phi {}_\phi \< x|$. 
In other words, by inspecting Eq. (\ref{otto}), we would like to write  an operator $G(x,\phi
)$ such that
\begin{equation}
\hbox{Tr}[G^\dag (x,\phi )a^{\dag n}a^m]=
{n+m \choose n}^{-1}
\,\frac {1}{\sqrt {2^{n+m}}} \,H_{n+m}(\sqrt 2 x)
\,e^{i\phi (m-n)}
\;.
\end{equation}
In this case, the operator $G(x,\phi ) $ is a dual of the
quadrature projectors $|x \> _\phi {}_\phi \< x|$, but is 
different from the canonical dual (\ref{6}), which is divergent for
unbounded operators. 
One has
\begin{equation}
\begin{split}
\dk{G^\dag (x, \phi )} 
&= \tilde F^{-1}\tilde F \dk{G^\dag (x, \phi )}=
\tilde F^{-1}  \sum_{k,l=0}^\infty 
\dk{ a^{\dag k}a^l }\db{ a^{\dag k}a^l}
\dk{G^\dag (x,\phi)} \\& = 
\sum_{k,l=0}^\infty 
\dk{g_{k,l}} \hbox{Tr}[a^{\dag l}a^k G^\dag (x,\phi)]\;.
\end{split}
\end{equation}
Then we obtain
\begin{equation}
G^\dag (x,\phi)= e^{i\phi a^\dag a } G^{\dag }(x,0) e^{-i\phi
  a^\dag a}\;,
\end{equation}
with 
\begin{equation}
\begin{split}
G^\dag (x,0)&= \sum_{k,l=0}^\infty 
g_{k,l} {k+l \choose k}^{-1}
\,\frac {1}{\sqrt {2^{k+l}}} \,H_{k+l}(\sqrt 2 x)
\\&= \sum_{k,l=0}^\infty \sum_{t=0}^{\min (k,l)}
\frac{(-1)^t}{t! \sqrt {k! l!}}a^t |k \> \< l| a^{\dag t}
{k+l \choose k}^{-1}
\,\frac {1}{\sqrt {2^{k+l}}} \,H_{k+l}(\sqrt 2 x)\\&= 
\sum_{k,l=0}^\infty \sum_{t=0}^{\min (k,l)}
\frac{(-1)^t}{t! \sqrt {(k-t)! (l-t)!}} |k -t\> \< l-t|\,
{k+l \choose k}^{-1}
\,\frac {1}{\sqrt {2^{k+l}}} \,H_{k+l}(\sqrt 2 x)
\;.
\end{split}
\end{equation}
The dual $G(x,\phi )$ provides also new
pattern functions for the matrix elements, as previously noticed in
Refs. \cite{Rich,Wun2}.  For example, for the vacuum
component one has 
\begin{equation}
\hbox{Tr}[G^\dag (x,\phi ) |0 \> \< 0|  ]=
\sum_{k=0}^\infty \left ( - \frac 12\right )^k \frac{k!}{(2k)!}
H_{2k}(\sqrt 2 x)\;.
\end{equation}
Notice that such pattern functions are generally no longer bounded for
$x \to \pm \infty $,  even for $\eta > 0.5$. 
\par One can check that the set $G(x,\phi )$ is dual to the 
set of quadrature projectors $|x \> _\phi {}_\phi \<
x|$ as follows
\begin{equation}
\begin{split}
&\int _{-\infty}^{+\infty }\d x \int_{0}^\pi \frac{\d\phi }{\pi } 
\dk{\delta (X_\phi -x)} \db{G(x,\phi )} \\& 
=\int _{-\infty}^{+\infty }\d x \int _{-\infty}^{+\infty }\d x' 
\int_{0}^\pi \frac{\d\phi }{\pi } 
\dk{|x' \> _\phi {}_\phi \< x'|} \db{G(x,\phi )}
\,\delta (x-x') \\ & 
=\int_{0}^\pi \frac{\d\phi }{\pi }\sum_{k,l=0}^\infty 
{k+l \choose k}^{-1}
\,\frac {1}{\sqrt {2^{k+l}}} 
e^{i\phi (k-l)}\,\dk{H_{k+l}(\sqrt 2 X_\phi )} \db{g^\dag
  _{k,l}}
\\ & 
=\sum_{k,l=0}^\infty \dk{a^{\dag l}a^k} \db{g^\dag _{k,l}} 
 =\sum_{k,l=0}^\infty \dk{a^{\dag l}a^k} \db{g_{l,k}}= \tilde F \tilde
 F^{-1}=I\otimes I\;.
\end{split}
\end{equation}
\section{Generating new frames}
We can generate different frames by changing the function of $a-b^\dag$ which gives the frame
operator in Eq. (\ref{4}). Explicitly, we have
\begin{equation}
\begin{split}
&f(|a-b^\dag|)=\int \frac {\d^2 z}{\pi} 
f(|z|) |D(z) \kk \bb D(z)|= \int _0^\pi \frac{\d\phi }{\pi }
\int _{-\infty}^{\infty} dt\, |t|\,f(|t|)\, |D(it e^{i\phi
}) \kk \bb D(ite^{i\phi })| \\ =& 
\int _0^\pi \frac{\d\phi }{\pi }\int
_{-\infty}^{\infty}\frac{\d x}{\pi}\, \int \d^2 z\int \d^2 w
|D(ze^{i\phi}) \kk \bb D(we^{i\phi }) |
\\  \times &  
g(|z|)g^*(|w|)  e^{2ix \hbox{\scriptsize Im}z - 2ix
\hbox{\scriptsize Im}w } \delta (\hbox{Re}z) \delta (\hbox{Re}w)\\
=&\pi \int _0^\pi \frac{\d\phi }{\pi }\int _{-\infty}^{\infty} \d x\, \int
\frac{\d^2 z }{\pi }\int \frac{\d^2 w}{ \pi} |D(z) \kk \bb
D(ze^{-i\phi})|g(|a^\dag -b|)| x \> _0 |x \> _0   
\\  \times &  
{}_0\< x |{}_0 \< x |g^*(|a^\dag-b|)| D(we^{-i\phi }) \kk \bb D(w)|\\
=&\int _0^\pi \frac{\d\phi }{\pi }\int _{-\infty}^{\infty} \d x\,
e^{i\phi (a^\dag a -b^\dag b)}\sqrt{\pi}g(|a^\dag-b|)| x \> _0 |x \> _0 {}_0\<
x |{}_0 \< x |\sqrt{\pi}g^*(|a^\dag-b|) e^{-i\phi (a^\dag a -b^\dag b)},
\end{split}
\end{equation}
where the functions $g$ and $f$ are related as follows
\begin{equation}
|t| f(|t|)=|g(t)|^2.
\end{equation}
Using the identity
\begin{equation}
(a-b^\dag)|A\kk=(|aA\kk-|Aa^*\kk)=|[a,A]\kk,
\end{equation}
for reference basis such that $a^*\equiv a$, e.~g. the photon number
basis, we have
\begin{equation}
|a-b^\dag|^2|A\kk=(a^\dag-b)|[a,A]\kk=|[a^\dag,[a,A]]\kk.
\end{equation}
But the double commutator can be written in terms of the (dual) Lindblad super-operator
\begin{equation}
[a^\dag,[a,A]]=-(\map{L}[a]+\map{L}[a^\dag])A,
\end{equation}
where $L[W]A\doteq W^\dag A W-\frac{1}{2}(W^\dag W A +AW^\dag W)$. Remarkably, this is exactly the
dissipative super-operator of the displacement Gaussian noise, corresponding to a distributed loss compensated
by a phase-insensitive amplification. \par With the aid of the following commutator rule
\begin{equation}
[a^\dag,[a,e^{i\lambda X}]]=\tfrac{1}{4}\lambda^2e^{i\lambda X}, 
\end{equation}
we easily obtain
\begin{equation}
|a^\dag -b|^2|x\>_0|x\>_0=|a -b^\dag|^2|x\>_0|x\>_0=-\tfrac{1}{4}\partial_x^2 
|\delta(x-X)\kk=|\mathcal{F}[\tfrac{1}{4}\lambda^2](X-x)\kk ,
\end{equation}
where $\mathcal{F}$ denotes the Fourier transform
\begin{equation}
\mathcal{F}[f](x)=\int_{-\infty}^\infty \frac{d \lambda}{2\pi}e^{i\lambda x}f(\lambda).
\end{equation}
Therefore, we have
\begin{equation}
g(|a^\dag -b|)|x\>_0|x\>_0=|\mathcal{F}[g(\tfrac{1}{2}\lambda)](X-x)\kk,
\end{equation}
corresponding to the frame
\begin{equation}
\Xi(x,\phi)=\sqrt{\pi}\mathcal{F}[g(\tfrac{1}{2}\bullet)](X_\phi-x).
\end{equation}
For example, if we choose $g(x)=\sqrt{\frac{\sigma^2}{2\pi}}\exp\left(-\frac{\sigma^2}{2}x^2\right)$, we
have
\begin{equation}
\Xi(x,\phi)=\sqrt{2\pi}\exp\left[-\frac{1}{\sigma^2}(X_\phi-x)^2\right].
\end{equation}
Notice that a function
\begin{equation}
\Xi(x,\phi)=h(X_\phi-x)
\end{equation}
corresponds to a frame if the function $h$ has Fourier transform which is invertible and
bounded. Moreover two functions $h$ and $h'$ will correspond to the same frame operator if their
Fourier transform have the same module. More precisely, the frame operator will be given by
\begin{equation}
F=f(|a-b^\dag|),\qquad f(t)=\frac{1}{\pi t
}\left|\mathcal{F}^{-1}[h](2t)\right|^2 \;.
\end{equation}
The canonical dual frame operator can be obtained by inverting the frame operator
\begin{equation}
\begin{split}
&F^{-1} | (|x \> _\phi {}_\phi \< x|) \kk =
\int\frac{\d^2 z}{\pi} \,\frac{1}{f(|z|)}|D(z) \kk \bb D(z)| x \> _\phi |x \>
_\phi \\ & =\int\frac{\d^2 z}{\pi} \,\frac{1}{f(|z|)} |D(ze^{i\phi}) \kk
e^{2i x \hbox{\scriptsize Im}z }\delta (\hbox{Re}z)  =
\int_{-\infty }^\infty \d k\,\frac{1}{f(|k|)}\,|D(-ike^{i\phi}) \kk e^{2ikx}
\\ &=\pi\int_{-\infty }^\infty \frac{d k}{2\pi}\,\frac{1}{f(|k|/2)}\,| e^{-ikX_\phi}\kk e^{ikx}
=|h^\vee(x-X_\phi)\kk,
\end{split}
\end{equation}
where
\begin{equation}
h^\vee(x)=\pi \mathcal{F}\left[\frac{1}{f(|k|/2)}\right](x).
\end{equation}
Hence, it follows that the usual Kernel operator corresponds to 
the canonical dual. 
\section{Other frames}
Using frame calculus, it is easy to show that the following sets of
operators are spanning sets
\begin{eqnarray}
&&A_{n,\phi }=e^{-\frac 12 X_\phi ^2}\,  X_\phi ^n
\nonumber \\& & 
B_{n,\phi }=\left( \frac{2}{\pi}\right )^{\frac 14}{\frac {1}{\sqrt {2^n
    n!}}}\,e^{-X_\phi ^2}H_n (\sqrt 2 X_\phi)
\;,
\end{eqnarray}
with corresponding frame operators 
\begin{eqnarray}
&&\int_{0}^\pi \frac{\d\phi }{\pi }\sum _{n=0}^\infty |A_{n,\phi} \kk
\bb A_{n,\phi}|=\frac{e^{-|Z|^2}}{|Z|}\;,
\nonumber \\& & 
\int_{0}^\pi \frac{\d\phi }{\pi }\sum _{n=0}^\infty |B_{n,\phi} \kk
\bb B_{n,\phi}|=\frac{1}{\pi |Z|}\;, 
\end{eqnarray}
where $Z=a-b^\dag$. 

\section*{Appendix}
\subsection{Proof of Lemma 4} 
Consider the Poisson form of the Dirac delta for the $2\pi$-interval
\begin{equation}
\delta_{2\pi}(\phi)=\lim_{\epsilon\to   1^-}\frac{1}{2\pi}\sum_{n=-\infty}^{+\infty}\epsilon^{|n|}e^{in\phi}.
\end{equation}
In the following we will use $\epsilon$ to mean $\epsilon=1_-$. 
Rescaling $\phi$ by a factor 2 we obtain
\begin{equation}
\delta_\pi(\phi)=\frac{1}{\pi}\sum_{n=-\infty}^{+\infty}\epsilon^{|n|}e^{i2n\phi}
\equiv\frac{1}{\pi}\sum_{n=-\infty}^{+\infty}\epsilon^{2|n|}e^{i2n\phi}\doteq\delta_\pi^{(+)}(\phi), 
\end{equation}
which is also equivalent to the {\em even folding} relation 
\begin{equation}
\delta_\pi^{(+)}(\phi)=\delta_{2\pi}(\phi)+\delta_{2\pi}(\phi+\pi),
\end{equation}
since
\begin{equation}
\begin{split}
\delta_{2\pi}(\phi)+\delta_{2\pi}(\phi+\pi)=&
\frac{1}{2\pi}\sum_{n=-\infty}^{+\infty}\epsilon^{|n|}\left[e^{in\phi}+(-)^ne^{in\phi}\right]\\
=&\frac{1}{\pi}\sum_{n=-\infty}^{+\infty}\epsilon^{2|n|}e^{i2n\phi}
\end{split}
\end{equation}
On the other hand, we have the {\em odd folding} relation
\begin{equation}
e^{i\phi}\delta_\pi^{(-)}(\phi)=\delta_{2\pi}(\phi)-\delta_{2\pi}(\phi+\pi)
\end{equation}
since
\begin{equation}
\begin{split}
&\delta_{2\pi}(\phi)-\delta_{2\pi}(\phi+\pi)=
\frac{1}{2\pi}\sum_{n=-\infty}^{+\infty}\epsilon^{|n|}\left[e^{in\phi}-(-)^ne^{in\phi}\right]\\
=&\frac{1}{\pi}\sum_{n=-\infty}^{+\infty}\epsilon^{|2n+1|}e^{i(2n+1)\phi}\doteq
e^{i\phi}\delta_\pi^{(-)}(\phi)
\end{split}
\end{equation}
From the equivalence relations (\ref{mainid}), we immediately derive the equivalence
\begin{equation}
\begin{split}\label{x2kd}
&x^{2k}\delta_\pi(\phi)\simeq\frac{1}{\pi} x^{2k}\sum_{n=-k}^k\epsilon^{|n|}e^{2in\phi}
=\frac{1}{\pi} x^{2k}\left[\frac{1-(\epsilon e^{2i\phi})^{k+1}}{1-\epsilon e^{2i\phi}}+
\frac{1-(\epsilon e^{-2i\phi})^{k+1}}{1-\epsilon e^{-2i\phi}}-1\right]\\
&=\frac{1}{\pi} x^{2k}\left[\frac{(\epsilon e^{2i\phi})^{k+1}}{\epsilon e^{2i\phi}-1}+
\frac{(\epsilon e^{-2i\phi})^{k+1}}{\epsilon e^{-2i\phi}-1}+\kappa_\epsilon(\phi)\right],
\end{split}
\end{equation}
where the distribution
\begin{equation}
\kappa_\epsilon(\phi)\doteq\frac{1-\epsilon^2}{1+\epsilon^2-\epsilon(e^{2i\phi}+e^{-2i\phi})}
\end{equation}
with support in $\phi=0$ gives
\begin{equation}
\begin{split}
&\int_0^\pi \frac{\d\phi}{\pi}\kappa_\epsilon(\phi)e^{i2n\phi}= 
\int_0^{2\pi}\frac{\d\phi}{2\pi}\frac{1-\epsilon^2}{1+\epsilon^2-\epsilon(e^{i\phi}+e^{-i\phi})}
e^{in\phi}\\=&\frac{1}{2\pi i}\oint\frac{\d z}{z}\frac{1-\epsilon^2}{(1-\epsilon z)(1-\epsilon z^{-1})}z^n=
\frac{1}{2\pi i}\oint\d
z\frac{\epsilon-\epsilon^{-1}}{(z-\epsilon^{-1})(z-\epsilon)}z^n= : \vartheta(n)
\end{split}
\end{equation}
where $\vartheta(n)=1$ for $n\ge 0$ and  $\vartheta(n)=0$ for $n< 0$, and the integral is performed
on the unit circle. Eq. (\ref{x2kd}), which contains identity (\ref{dicomb}) as a special case,
generalizes as follows
\begin{equation}
f(x^2)\delta_\pi(\phi)\simeq 
\frac{1}{\pi}\left[\frac{f(\epsilon x^{2}e^{2i\phi})}{1-\epsilon^{-1}e^{-2i\phi}}+
\frac{f(\epsilon x^{2}e^{-2i\phi})}{1-\epsilon^{-1}e^{2i\phi}}+\kappa_\epsilon(\phi)f(x^2)\right],
\end{equation}
where $f(z)$ denotes any analytic function in $z$. In a similar way we calculate
\begin{equation}
\begin{split}
&x^{2k+1}e^{i\phi}\delta_\pi(\phi)\simeq\frac{1}{\pi} x^{2k+1}\sum_{n=-k-1}^k\epsilon^{|n|}
e^{i(2n+1)\phi}\\
=&\frac{1}{\pi} x^{2k+1}e^{i\phi}\left[\frac{1-(\epsilon e^{2i\phi})^{k+1}}{1-\epsilon e^{2i\phi}}+
\frac{1-(\epsilon e^{-2i\phi})^{k+2}}{1-\epsilon e^{-2i\phi}}-1\right]\\
&=\frac{1}{\pi} x^{2k+1}e^{i\phi}\left[\frac{(\epsilon e^{2i\phi})^{k+1}}{\epsilon e^{2i\phi}-1}+
\frac{(\epsilon e^{-2i\phi})^{k+2}}{\epsilon e^{-2i\phi}-1}+\kappa_\epsilon(\phi)\right],
\end{split}
\end{equation}
which generalizes as follows
\begin{equation}
xf(x^2)e^{i\phi}\delta_\pi(\phi)\simeq \frac{1}{\pi}x
\left[\frac{e^{i\phi}f(\epsilon x^2e^{2i\phi})}{1-\epsilon^{-1} e^{-2i\phi}}+\frac{e^{-i\phi}
f(\epsilon x^2e^{-2i\phi})}{1-\epsilon^{-1}e^{2i\phi}}+\kappa_\epsilon(\phi)e^{i\phi}f(x^2)
\right].
\end{equation}
\subsection{Alternative derivation of identity (\ref{estr})}
By posing $\alpha=\frac{i}{2}re^{i\phi}$, we have
\begin{equation}
\begin{split}
D(\alpha)=&\int_{-\pi}^\pi \frac{\d\phi}{2\pi}2\pi\delta_{2\pi}(\phi-\phi) e^{ir X_\phi}\\
=&\int_0^\pi \frac{\d\phi}{\pi}\pi\left[\delta_{2\pi}(\phi-\phi)e^{ir X_\phi}+\delta_{2\pi}
(\phi-\phi-\pi)e^{-ir X_\phi}\right]\\
=&\int_0^\pi \frac{\d\phi}{\pi}\pi\left[\cos rX_\phi(\delta_{2\pi}(\phi-\phi)+\delta_{2\pi}
(\phi-\phi-\pi))\right.\\
+&\left.i\sin rX_\phi(\delta_{2\pi}(\phi-\phi)-\delta_{2\pi}(\phi-\phi-\pi))\right]\\
=&\int_0^\pi \frac{\d\phi}{\pi}\pi\left[\cos rX_\phi\delta_\pi^{(+)}(\phi-\phi)
+i\sin rX_\phi\delta_\pi^{(-)}(\phi-\phi)\epsilon e^{i(\phi-\phi)}\right].
\end{split}
\end{equation}
Now, we evaluate separately the cosine and the sine terms. In the following, we will denote
$\psi=\phi-\phi$. The cosine term can be transformed as follows
\begin{equation}
\begin{split}
&\cos rX_\phi\pi\delta_\pi^{(+)}(\psi)
\simeq\sum_{k=0}^\infty \frac{(-)^k}{(2k)!}r^{2k}X_\phi^{2k}\sum_{n=-k}^k  
\epsilon^{2|n|}e^{2i n\psi}\\=&\sum_{k=0}^\infty \frac{(-)^k}{(2k)!}r^{2k}X_\phi^{2k}
\left[\sum_{n=0}^k\epsilon^{2n}\left(e^{2in\psi}+e^{-2in\psi}\right)-1\right]\\
=&\sum_{n=0}^\infty\epsilon^{2n}\left(e^{2in\psi}+e^{-2in\psi}\right)
\sum_{k=n}^\infty \frac{(-)^k}{(2k)!}r^{2k}X_\phi^{2k}-\cos(rX_\phi)\\
=&\sum_{n=0}^\infty\epsilon^{2n}\left(e^{2in\psi}+e^{-2in\psi}-\delta_{n0}\right)
(-)^nr^{2n}X_\phi^{2n}\sum_{k=0}^\infty \frac{(-)^kr^{2k}X_\phi^{2k}}{(2k+2n)!}
\end{split}
\end{equation}
On the other hand, the sine term transforms as follows
\begin{equation}
\begin{split}
&\sin rX_\phi\pi\delta_\pi^{(-)}(\psi)\epsilon e^{i\psi}\simeq\sum_{k=0}^\infty \frac{(-)^kr^{2k+1}X_\phi^{2k+1}}{
(2k+1)!}\sum_{n=-k-1}^k\epsilon^{|2n+1|}e^{i(2n+1)\psi}\\=&
\sum_{k=0}^\infty \frac{(-)^kr^{2k+1}X_\phi^{2k+1}}{(2k+1)!}\left(\sum_{n=0}^k\epsilon^{2n+1}
e^{i(2n+1)\psi}+\sum_{n=1}^{k+1}\epsilon^{2n-1} e^{-i(2n-1)\psi}\right)\\=&
\sum_{k=0}^\infty \frac{(-)^kr^{2k+1}X_\phi^{2k+1}}{(2k+1)!}\left(\sum_{n=0}^k\epsilon^{2n+1}
e^{i(2n+1)\psi}+\sum_{n=0}^k\epsilon^{2n+1} e^{-i(2n+1)\psi}\right)\\=&
\sum_{n=0}^\infty\epsilon^{2n+1}\left(e^{i(2n+1)\psi}+e^{-i(2n+1)\psi}\right)
\sum_{k=n}^\infty\frac{(-)^kr^{2k+1}X_\phi^{2k+1}}{(2k+1)!}\\=&
\sum_{n=0}^\infty\epsilon^{2n+1}\left(e^{i(2n+1)\psi}+e^{-i(2n+1)\psi}\right)r^{2n}X_\phi^{2n}
\sum_{k=0}^\infty\frac{(-)^kr^{2k+1}X_\phi^{2k+1}}{(2k+2n+1)!}
\end{split}
\end{equation}
It is convenient now to make the following substitutions
\begin{equation}
r=2|\alpha|,\qquad e^{-i\phi }=i \frac{\alpha^*}{|\alpha|},\qquad e^{i\psi}=  
i\frac{\alpha^*}{|\alpha|}e^{i\phi},\qquad e^{i\psi}r =2i\alpha e^{i\phi}.
\end{equation}
In this way, the cosine term becomes
\begin{equation}
\begin{split}
&\cos rX_\phi\pi\delta_\pi^{(+)}(\psi)\\ \simeq  &
\sum_{n=0}^\infty\left((\epsilon\alpha^*)^{2n}e^{i2n\phi}+(\epsilon\alpha)^{2n}e^{-i2n\phi}-\delta_{n0}
\right)(2X_\phi)^{2n}\sum_{k=0}^\infty \frac{(-)^k(2|\alpha|X_\phi)^{2k}}{(2k+2n)!},
\end{split}
\end{equation}
and the sine term 
\begin{equation}
\begin{split}
&i\sin rX_\phi\pi\delta_\pi^{(-)}(\psi)\epsilon e^{i\psi}\\ \simeq &
\sum_{n=0}^\infty\left(-(\epsilon\alpha^*)^{2n+1}e^{i(2n+1)\phi}+(\epsilon\alpha)^{2n+1}e^{-i(2n+1)\phi}
\right)(2X_\phi)^{2n+1}\sum_{k=0}^\infty\frac{(-)^k(2|\alpha|X_\phi)^{2k}}{(2k+2n+1)!},
\end{split}
\end{equation}
and putting the two terms together we have
\begin{equation}
D(\alpha)=\int_0^\pi \frac{\d\phi}{\pi}f_\phi^\epsilon(X_\phi|D(\alpha)),
\end{equation}
with
\begin{equation}
\begin{split}
&f_\phi^\epsilon(X_\phi|D(\alpha))=\sum_{n=0}^\infty 
\left((-\epsilon\alpha^*)^n e^{in\phi}+(\epsilon\alpha)^ne^{-in\phi}-\delta_{n0}
\right)(2X_\phi)^n\sum_{k=0}^\infty\frac{(-)^k(2|\alpha|X_\phi)^{2k}}{(2k+n)!}\\
=& \sum_{n=0}^\infty\sum_{k=0}^\infty
\left(\epsilon^ne^{in\phi}(-\alpha^*)^{n+k}\alpha^k +\epsilon^ne^{-in\phi}
(-\alpha^*)^k\alpha^{n+k}-\delta_{n0}
\right)\frac{(2 X_\phi)^{2k+n}}{(2k+n)!}
\end{split}
\end{equation}
In the limit $\epsilon\to 1^-$ the last expression can be simplified using the reordering rule
\begin{equation}
\begin{split}
&\sum_{n=0}^\infty\sum_{k=0}^\infty a_{n+2k}(z^{n+k}w^kt^n+ z^kw^{n+k}t^{-n}-\delta_{n0}z^kw^k)\\
=&\sum_{k=0}^\infty\sum_{h=k}^\infty a_{h+k}(z^hw^kt^{h-k}+ z^kw^ht^{k-h}-\delta_{hk}z^kw^k)\\
=&\sum_{h=0}^\infty\sum_{k=0}^h a_{h+k}(tz)^h(t^{-1}w)^k+\sum_{k=0}^\infty\sum_{h=k}^\infty 
a_{h+k}(tz)^k(t^{-1}w)^h-\sum_{h=0}^\infty a_{2h}z^hw^h\\
=&\sum_{h=0}^\infty\sum_{k=0}^\infty a_{h+k}(tz)^h(t^{-1}w)^k,
\end{split}
\end{equation}
corresponding to
\begin{equation}
f_\phi(X_\phi|D(\alpha))=\sum_{h=0}^\infty\sum_{k=0}^\infty
\frac{(-2 X_\phi e^{i\phi}\alpha^*)^h(2 X_\phi e^{-i\phi}\alpha)^k}{(h+k)!}\;.
\end{equation}

\bibliographystyle{unsrt
}\bibliography{/Users/dariano/BOOK/book}

\end{document}